\newcommand{\bfl}{\begin{flushleft}}
\newcommand{\efl}{\end{flushleft}}
\newcommand{\bc}{\begin{center}}
\newcommand{\ec}{\end{center}}
\newcommand\ie {{\it i.e. }}

\newcommand\emptypage{~~~ \eject}
\def\be{\begin{eqnarray}}
\def\ee{\end{eqnarray}}
\newenvironment{draftequation}[1]{\be\label{#1}}{\ee}
\newcommand\bbe[1]{\begin{draftequation}{#1}}
\newcommand\eee{\end{draftequation}}
\def\half{{\textstyle{1 \over 2}}}

\newcommand\art[1]{\cite{#1}}
\newcommand\ekv[1]{(\ref{#1})}

\documentstyle[preprint,aps,psfig,here]{revtex}

\begin{document}

\draft


\title{
\begin{flushright}
{\rm Oslo-TP 6-95,
USITP-95-05\\
gr-qc/9504027}
\end{flushright}
 Rotating Dilaton Black Holes}


\author{Bj\o rn Jensen\footnote{Electronic address: BJensen@boson.uio.no}}

\address{
Institute of Physics, University of Oslo,
P.O. Box 1048, N-0316 Blindern, Oslo 3, Norway}

\author{Ulf Lindstr\"{o}m\footnote{Electronic address: UL@vanosf.physto.se and
ulfl@boson.uio.no}}

\address{ITP, University of Stockholm,
Box 6730, S-113 85, Stockholm, Sweden }

\date{\today}

\bibliographystyle{unsrt}

\maketitle

\begin{center}
{\em Manuscript submitted to Phys.Rev.D: Rapid Communication}
\end{center}


\begin{abstract}
We consider the axially symmetric coupled system of gravitation,
electromagnetism and a dilaton field. Reducing from four to three dimensions,
the system is described by gravity coupled to a non-linear $\sigma$-model. We
find the target space isometries and use them to generate new solutions. It
seems that it is only possible to generate rotating solutions from non-rotating
ones for the special cases when the dilaton coupling parameter $a=0, \pm
\sqrt{3}$. For those particular values, the target space symmetry is enlarged.
\end{abstract}

\emptypage

\bfl
{\bf Introduction}\\
\efl
\bigskip

Dilaton Black holes, \ie black hole solutions with a scalar field
(dilaton) and electromagnetic fields present,
 show qualitatively different behavior
as compared to solutions in pure Einstein-Maxwell gravity
\art{HOWI,BJUL1}.
In this letter we report on new classes of rotating
dilaton black hole solutions. We
derive these solutions using a generating technique that goes back to
Neugebauer, Kramer and
Geroch \art{Kramer,GERO}. This technique was recently used in
\art{GAKE} to generate
axion/dilaton solutions. One result of our investigation is that the space
of solutions becomes drastically reduced when the axion is not present. In
fact, using our approach it seems possible to generate rotating solutions
from  non-rotating ones only for certain values of the dilaton coupling
parameter.

\bigskip
\bfl
{\bf The $\sigma$ model action}\\
\efl
\bigskip

We consider the following action:
\bbe{act1}
S=\frac{1}{16\pi}\int \sqrt{-g}d^4x(-^{(4)}R+2\partial_\mu\phi\partial^\mu\phi
-e^{-2a\phi}F_{\mu\nu}F^{\mu\nu})
\eee
where $^{(4)}R$ is the curvature scalar formed from the
space-time metric $g_{\mu
\nu}$, $\phi$ is the dilaton and
$F_{\mu\nu}$ is the electromagnetic field strength. The coupling of the dilaton
is governed by the real
parameter $a$ \art{HOWI}. The most general four-dimensional
axi-symmetric metric may be written:
\bbe{axmet}
^{(4)}ds^2=f(dt-\omega_idx^i )^2-f^{-1}h_{ij}dx^idx^j\, ,
\eee
where $h_{ij}$,
$f$ and $\omega_i$ are independent of $t$. As suggested
by \ekv{axmet}, we focus on a three dimensional hypersurface embedded in four
space. We rewrite the action
(\ref{act1}) in terms of the intrinsic and extrinsic curvatures,
$^{(3)}R_{ij}$ and $K_{ij}$, as \art{WALD}:
\bbe{act2}
S=\int d^3x N|h|^{1/2}(\, ^{(3)}R(\tilde h_{ij})-K_{ij}K^{ij}+K^2)\, .
\eee
where  $N$ is the laps function and $\tilde h_{ij} = -f^{-1}h_{ij}$ is the
metric on the hypersurface perpendicular to the normal vector
field $\vec{n}$ defined by
$n_\mu dx^\mu =f^{1/2}(dt-\omega_i dx^i)$. We further define the twist
potential $\chi$ trough
\bbe{chi}
\partial_{[j}\omega_{k]}=
-(\sqrt{h}f^2)^{-1}\epsilon^i\, _{jk}\partial_i\chi
\eee
and the electric and magnetic potentials $v$ and $u$ through
\bbe{uv}
F_{i0}\equiv\frac{1}{\sqrt{2}}\partial_i v\,\, ,\,\,
F^{ij}\equiv\frac{1}{\sqrt{2h}}e^{2a\phi}f^3\epsilon^{ijk}
\partial_k u\, ,
\eee
where indices now are raised and lowered by the metric $h_{ij}$.
After a rescaling of the 3-metric $\tilde h_{ij}$ in \ekv{act2}, we
may write the the total action \ekv{act1} as
the action for a  3D nonlinear $\sigma$-model
coupled to gravity:
\bbe{sigm}
S=\int ( ^{(3)}R({h})
-G_{AB}\partial_{i}\Phi^A\partial^i\Phi^B)\sqrt{h}d^3x\, .
\eee
The $\sigma$-model target space is coordinatized by the
fields $\{\Phi ^A\} \equiv \{\phi ,f,\chi ,u,v,\}$ and the target space metric
can be read
off from the line-element
\bbe{line}
dl^2=2d\phi^2+\frac{1}{2f^2}(df^2+(d\chi +vdu-udv)^2)-\frac{1}{f}(e^{-2a\phi}
dv^2+e^{2a\phi}du^2)\, .
\eee

\bigskip
\bfl
{\bf Isometries}\\
\efl
\bigskip

The action \ekv{sigm} and the corresponding field equations are invariant under
target space isometries, \ie transformations of $\Phi ^A$ that leave \ekv{line}
invariant. This fact may be utilized to generate new solutions from known ones.
To do so we must first find the isometries, \ie find the Killing vectors that
generate the infinitesimal isometries and
then exponentiate to find the finite
transformations.

Letting $\vec{\xi}^\alpha$ denote the Killing vectors (denumbered by $\alpha$),
 Killings equation reads
\bbe{Kill}
\xi^\alpha _{(B:C)}=0,
\eee
where $:$ denotes (target space) covariant derivative. The solutions to
\ekv{Kill} differ depending on the value of the dilaton coupling parameter
$a$. For arbitrary values of $a$, we find four Killing vectors:
\bbe{1-4}
\vec{\xi}^1&=&2f\partial_f+2\chi\partial_\chi +u\partial_u+v\partial_v\cr
\vec{\xi}^2&=&v\partial_\chi +\partial_u\cr
\vec{\xi}^3&=&u\partial_\chi -\partial_v\cr
\vec{\xi}^4&=&\partial_\chi \,.
\eee
A fifth Killing vector,
\bbe{5a}
\vec{\xi}^5&=&\frac{1}{a}\partial_\phi -u\partial_u+v\partial_v\, .
\eee
is only defined for $a\ne 0$. When $a=0$ we find instead
\bbe{50}
\hat{\vec{\xi}^5}=v\partial_u-u\partial_v\, .
\eee
The vectors ${\bf \xi}^1,...,{\bf \xi}^5$ form a closed algebra, and for
generic
$a$'s, these
 exhaust the solutions. For the special values $a=0$ and $a^2=3$ there are
more solutions, however. For $a=0$ we find the additional solutions
\bbe{6-9}
\hat{\vec{\xi}^6}&=&\partial_\phi\cr
\hat{\vec{\xi}^7}&=&2fu\partial_f+(u\chi
+\frac{1}{2}v(v^2+u^2)-vf)\partial_\chi
+\cr &+&
\frac{1}{2}(u^2-3v^2+2f)\partial_u+(2uv-\chi )\partial_v\cr
\hat{\vec{\xi}^8}&=&2fv\partial_f+(v\chi
-\frac{1}{2}u(v^2+u^2)+uf)\partial_\chi+\cr &+&
\frac{1}{2}(v^2-3u^2+2f)\partial_v+(2uv+\chi )\partial_u\cr
\hat{\vec{\xi}^9}&=&-4f\chi\partial_f+(-2f(u^2+v^2)+\frac{1}{2}(u^2+v^2)^2+2(f^2-\chi
^2))\partial_\chi
-\cr&-&(2u\chi +u^2v+v^3-2fv)\partial_u+(-2v\chi +uv^2+u^3-2fu)\partial_v\, .
\eee
These vectors also form a closed algebra, where $\hat{\vec{\xi}^6}$ commutes
with all generators and $\hat{\vec{\xi}^7},...,\hat{\vec{\xi}^9}$ form a
sub-algebra. When
$\phi =0$, $\hat{\vec{\xi}^7}$ and $\hat{\vec{\xi}^8}$ generate Harrison
type transformations
and commute to $\hat{\vec{\xi}^9}$ which generates Ehlers type
transformations
\art{EHLE}.

Finally and most interestingly, for $a^2=3$ we have the following additional
set
of Killing vectors (for definiteness we choose $a=\sqrt{3}$):
\bbe{7-8}
\tilde{\vec{\xi}^6}&=&-\sqrt{3}u\partial_\phi +2fu\partial_f+(u(\chi -uv)-e^{-
2\sqrt{3}\phi}fv)\partial_\chi +\cr&+&(2u^2+e^{-2\sqrt{3}\phi}f)\partial_u-
(\chi +uv)\partial_v\cr
\tilde{\vec{\xi}^7}&=&\sqrt{3}v\partial_\phi +2fv\partial_f +(v(\chi
+uv)+e^{+2\sqrt {3}\phi}fu)\partial_\chi\cr&+&
(2v^2+e^{+2\sqrt{3}\phi}f)\partial_v+(\chi -uv)\partial_u\cr
\tilde{\vec{\xi}^8}&=&\sqrt{3}uv\partial_\phi +2\chi
f\partial_f+(e^{2\sqrt{3}\phi}fu^ 2+
e^{-2\sqrt{3}\phi}fv^2+u^2v^2+\chi^2-f^2)\partial_\chi-\cr&-&(fve^{-2\sqrt{3}\phi
}+u^2v-u\chi )\partial_u+(fue^{2\sqrt{3}\phi}+uv^2+v\chi )\partial_v\, .
\eee
For this case the full multiplication table is given in fig.1.
In this table, we again find a subalgebra generated by the last three vectors
$\tilde{\vec{\xi}^6},...,\tilde{\vec {\xi}^8}$. Here $\tilde{\vec{\xi}^6}$ and
$\tilde{\vec{\xi}^7}$ generate (dual) Harrison type transformations and
$\tilde{\vec{\xi}^8}$ generates an Ehlers type transformation.

\bigskip
\bc
{\bf The finite transformations}
\ec

\bigskip

To find the finite transformations, we exponentiate the generators
\ekv{1-4}, \ekv{5a},\ekv{50},\ekv{6-9} and \ekv{7-8} multiplied by the
corresponding group parameters $\lambda_\alpha$. An isometry then acts on the
fields as follows:
\bbe{fin}
{\Phi '} ^A=exp\{\lambda_\alpha \vec \xi^\alpha\}\Phi ^A.
\eee
For a general $a \ne 0$ we find from \ekv{1-4},\ekv{5a}:
\bbe{fin1}
n=1&:&\phi '=\phi\, ,\, f'=e^{2\lambda_1}f\, ,\, \chi '=e^{2\lambda_1}\chi\,
,\,
u'=e^{\lambda_1}u\, ,\, v'=e^{\lambda_1}v\cr
n=2&:&\phi'=\phi\, ,\, f'=f\, ,\, v'=v\, ,\, \chi '=\chi +v\lambda_2\, ,\,
u'=u+\lambda_2\cr
n=3&:&\phi' =\phi\, ,\, f'=f\, ,\, u'=u\, ,\, \chi '=\chi +u\lambda_3\, ,\,
v'=v-\lambda_3\cr
n=4&:&\phi'=\phi\, ,\, f=f'\, ,\, u'=u\, ,\, v'=v\, ,\, \chi '=\chi
+\lambda_4\cr n=5&:&f'=f\, ,\, \chi '=\chi\, ,\, \phi '=\phi +a^{-1}\lambda_5\,
,\, u'=e^{-\lambda_5}u\, ,\, v'=e^{\lambda_5}v\, .
\eee
For $a=0$ two additional transformations are:
\bbe{fin2}
n=5&:&\phi '=\phi\, ,\, f'=f\, ,\, \chi '=\chi\, ,\, u'=\cos (\hat \lambda_5)u+
\sin (\hat \lambda_5)v\, ,\, v'=\cos (\hat \lambda_5)v-\sin (\hat \lambda_5)u
,\cr  n=6&:& \phi'=\phi +\hat \lambda_6\, ,\, f=f'\, ,\, u'=u\, ,\, v'=v\, ,\,
\chi '=\chi\, .
\eee
The finite transformations that correspond to $\tilde{\vec
{\xi}^7},...,\tilde{\vec
{\xi}^9}$ in \ekv{6-9} are more complicated. We do not give them here, since
the
minimal coupling of a dilaton to gravity and electromagnetism
reduces to Einstein-Maxwell theory which has been studied in our
type of approach in \art{Kramer}.

For $a^2=3$ the finite transformations are \ekv{fin2} above with $a=\pm \sqrt
3$ along with those that correspond to $\tilde{\vec
{\xi}^6},...,\tilde{\vec{\xi}^8}$ in
\ekv{7-8}. From the multiplication table in Fig.1 we see that
$\tilde{\vec{\xi}^8}$ commutes with all other generators. This means that the
corresponding group element is known once the finite transformations
generated by $\tilde{\vec{\xi}^6}$ and $\tilde{\vec{\xi}^7}$ have been found:
\bbe{F8}
exp\{(\tilde\lambda_8 )^2\tilde{\vec {\xi}^8}\}=exp\{\tilde\lambda_8
\tilde{\vec {\xi}^6}\}exp\{\tilde\lambda_8 \tilde{\vec {\xi}^7}\}
exp\{-\tilde\lambda_8
\tilde{\vec {\xi}^6}\}exp\{-\tilde\lambda_8 \tilde{\vec {\xi}^7}\}.
\eee
The finite transformations that follow from $\tilde{\vec {\xi}^6}$ are
\bbe{F6}
U'^{-1}_\pm=U^{-1}_\pm -\lambda _6\quad
v'=v-\Xi\lambda_6\quad
F'=F,\quad \Xi'=\Xi ,
\eee
where (for $a=+\sqrt{3}$)
\bbe{deF6}
F\equiv exp\{2\phi /\sqrt{3}\}f, \quad \Xi\equiv \chi + uv\quad
U_\pm\equiv 2u\pm\sqrt{2F}exp\{-4\phi /\sqrt{3}\}.
\eee
The finite transformations generated by $\tilde{\vec {\xi}^7}$ are very
similar,
due to the "duality" symmetry $u\rightarrow v\, ,\,
v\rightarrow -u\, ,\, \phi\rightarrow -\phi$. They are
\bbe{F7}
V'^{-1}_\pm=V^{-1}_\pm -\lambda _7\quad
u'=u+\Xi\lambda_7\quad
F'=F,\quad \Xi'=\Xi ,
\eee
where
\bbe{deF7}
F\equiv exp\{-2\phi /\sqrt{3}\}f, \quad \Xi\equiv \chi - uv\quad
V_\pm\equiv 2v\pm\sqrt{2F}exp\{4\phi /\sqrt{3}\}.
\eee

\newpage

\bfl
{\bf New Solutions}
\efl

\bigskip

We are now in position to generate new solutions from old ones using the
isometry transformations just derived. The technique is to start from a known
seed solution and then apply one or several transformations
to it. It turns out
that the transformations \ekv{fin1} and \ekv{fin2} do not lead to
interesting new solutions in general. We will thus focus on the Harrison (and
Ehlers)-type transformations \ekv{F8}, \ekv{F6}, \ekv{F7}. Below we briefly
present several examples. We will give a fuller description elsewhere
\art{BJULINPREP}.

\bigskip

\bfl
{\bf Example 1.}\\
\efl

We first use a general vacuum solution as seed solution: $u_0=v_0=\phi_0=0$.
The original field variables are thus $f_0$, and $\chi_0$. Acting on these with
\ekv{F6} and \ekv{F7} we find
\bbe{X1}
\phi &=& {\sqrt{3}\over 4}ln(1-2f_0\lambda ^2), \quad
f=f_0/\sqrt{1-2f_0\lambda ^2}\quad
\chi = \left({1-f_0\lambda ^2}\over{1-2f_0\lambda ^2}\right)\chi_0,\cr
u&=&\lambda \chi_0, \quad
v={{f_0\lambda}\over{1-2f_0\lambda ^2}}.
\eee
The transformation generated by \ekv{F7} result if we make the substitution
$u\rightarrow v\, ,\, v\rightarrow -u\, ,\, \phi\rightarrow -\phi$. If
the original solution has fields that go to zero at infinity, so will the new
one. It is also clear that whereas we generate matter fields using these
transformations, we cannot generate a rotating solution from a non-rotating
one.

\bigskip

\bfl
{\bf Example 2.}\\
\efl

As a second seed solution, we use the static dilaton black-hole solution
\art{HOWI}, for $a=\sqrt{3}$:
\bbe{X2}
f_0={\Delta\over{\sigma^2}},\quad exp\{2\sqrt{3}\phi_0\}=\sigma^2,\quad
v_0={Q\over r},\quad u_0=0,\quad \chi _0=0,
\eee
where
\bbe{X2DEF}
\Delta\equiv \left(1-{{r_+}\over r}\right)\quad \sigma^2\equiv
\left(1-{{r_-}\over r}\right)^{3\over 2}\quad
r_+r_-=2Q^2,\quad -2(2M-r_+)=r_-,
\eee
with $M$ and $Q$ the mass and charge of the black hole, respectively.
Applying the transformations \ekv{F7} to the fields in \ekv{X2} we obtain
\bbe{RX2}
exp\{2\sqrt{3}\phi\}&=&\sigma^2\left[\left(1-{{2\lambda Q}\over
r}\right)^2-2\lambda\Delta\right]^{-{{3}\over 2}},\quad
f=\Delta (\sigma^2)^{-1}\left[\left(1-{{2\lambda Q}\over
r}\right)^2-2\lambda\Delta\right]^{-{1\over 2}}, \cr
\chi &=&0, \quad u=0,\quad
v=\left[\left(1-{{2\lambda Q}\over
r}\right)^2-2\lambda\Delta\right]^{-1}\left({Q\over r}\left(1-{{2\lambda
Q}\over
r}\right)+\lambda\Delta\right).
\eee
Again, no rotation. However, the seed solution is asymmetric in $u$ and $v$,
and will therefore be treated differently by the two transformations \ekv{F6}
and \ekv{F7}. In fact, the result of \ekv{F6} is:
\bbe{R2X2}
&exp&\{2\sqrt{3}\phi_1\}=
(\sigma^2)^{-2}\left((\sigma)^2-2\lambda^2\Delta\right)^{3\over 2}\quad
f_1=\Delta\left((\sigma)^2-2\lambda^2\Delta\right)^{-{1\over 2}}, \cr
&\chi_1& =
-{{Q\lambda\Delta}\over
r}\left((\sigma)^2-2\lambda^2\Delta\right)^{-1},\quad
u_1=\Delta\lambda\left((\sigma)^2-2\lambda^2\Delta\right)^{-1},\quad
v_1={Q\over
r}.
\eee
We see that this produces a rotating solution. When $r\to \infty$ the
rotation goes to zero while $f$ and $u$ go to constant values. After a
rescaling generated by $\vec\xi_1$ with parameter $\half
ln(\sqrt{1-2\lambda^2})$
and a $\vec\xi_2$-transformation with parameter $-\lambda/(1-2\lambda^2)^{3/4}$
we obtain the following asymptotically flat rotating dilaton black-hole
solution:
\bbe{RBH}
\phi &=&\phi_1, \quad f=Af_1, \quad  ,
u=A^{1\over 2} u_1-A^{{-3}\over 2}\lambda,\quad v=A^{1\over 2} v_1\cr
\chi &=&A\chi_1 -A^{-1}\lambda v_1 \quad \Rightarrow \omega_\theta=\omega_r=0,
\quad \omega_\phi = \left(2\lambda Q A^{-1}\right)cos\theta,
\eee
where $A\equiv \sqrt{1-2\lambda ^2}$. This is a Taub-NUT type solution.
That type of a solution has also been found
in dilaton-axion gravity for $a=1$ \cite{Kallosh}.

\bigskip

\bfl
{\bf Example 3.}\\
\efl

As our final example, we apply the formulae in \ekv{X1} using the Kerr solution
as a seed solution. The starting point is thus
\bbe{Kerr}
\phi_0 = 0, \quad f_0={{\Delta - \rho^2sin^2\theta}\over \Sigma}\quad \chi_0 =
{{2M\rho cos\theta}\over \Sigma},\quad u_0=v_0=0,
\eee
where the coordinates are $t,r,\varphi,\theta$, and
\bbe{KerrD}
\Delta \equiv r(r-2M)+\rho^2,\quad \Sigma\equiv r^2+\rho^2cos^2\theta .
\eee
Here $M$ is the mass and $\rho$ is the angular momentum parameter. The
(diagonal) three-metric is
\bbe{3met}
h_{rr}={{\Delta - \rho^2sin^2\theta}\over \Delta}, \quad h_{\theta \theta}
=\Delta - \rho^2sin^2\theta ,\quad h_{\varphi\varphi}=\Delta sin^2\theta .
\eee
{}From \ekv{X1} we find
\bbe{frol}
exp\{2\sqrt{3}\phi_1\}=\left[1-{{\Delta - \rho^2sin^2\theta}\over
\Sigma}\right]^{-{3\over 2}}\quad f_1={{\Delta - \rho^2sin^2\theta}\over
{\sqrt{\Sigma^2-2\lambda^2\Sigma{\Delta - \rho^2sin^2\theta}}}},\cr
\nonumber\\
\chi_1={{2M\rho cos\theta ({\Sigma^2-\lambda^2\Sigma{\Delta -
\rho^2sin^2\theta}})\over{\Sigma ({\Sigma^2-2\lambda^2\Sigma{\Delta -
\rho^2sin^2\theta}})}}},\quad u_1={{2M\rho \lambda cos\theta}\over\Sigma},\quad
v_1={{\lambda (\Delta - \rho^2sin^2\theta )}\over {\Sigma -2\lambda^2({\Delta -
\rho^2sin^2\theta})}}.\nonumber
\eee
A $\vec\xi_1$ rescaling with parameter $\half ln(A)$ and a $\vec\xi_3$
transformation with parameter $\lambda /A^{3\over 2}$, ($A$ as in example 2),
lead to the final asymptotically flat result
\bbe{Rfro}
\phi = \phi_1,\quad f=Af_1,\quad \chi =A\chi_1 +\lambda A^{-1}u_1,\quad
u=A^{1\over 2}u_1, \quad v=A^{1\over 2}v_1-\lambda A^{-3\over 2}.
\eee
This is the rotating dilaton black hole solution found in \art{FROL}

\bigskip

\bfl
{\bf Discussion}
\efl

Since the transformations \ekv{fin1} give trivial
 new solutions from
vacuum solutions and they exhaust the symmetries for generic $a$'s, rotating
dilaton black hole solutions seem harder to come by than the axion-dilaton ones
\art{GAKE}. It is clear from our discussion that there are families of rotating
solutions for $a=0, \pm \sqrt{3}$, however. We have given some
examples, and
more examples are easily available. One may use e.g
the solutions \ekv{R2X2} and \ekv{Rfro}
as seed solutions. The special values of $a$ correspond to minimal
couplings of the dilaton field. For $a=0$ the coupling to the $F^2$ term in the
lagrangian is 1 in $D=4$. The value $a^2=3$ corresponds to minimal coupling in
$D=5$, and
it has previously been encountered in many cases as resulting from dimensional
reduction \art{FROL,Gibbons2}.\\

\bigskip
{\bf Acknowledgement}: The research of UL was supported in part by
NFR grant No F-AA/FU04038-312 and by NorfA grant No. 94.35.049-0.
BJ thanks the University of Stockholm for hospitality during the time
in which parts of this work was carried out, and NFR for a travelling grant,
grant no. 420.94/013.

\begin{figure}[htbp]

\begin{center}

\mbox{\psfig{file=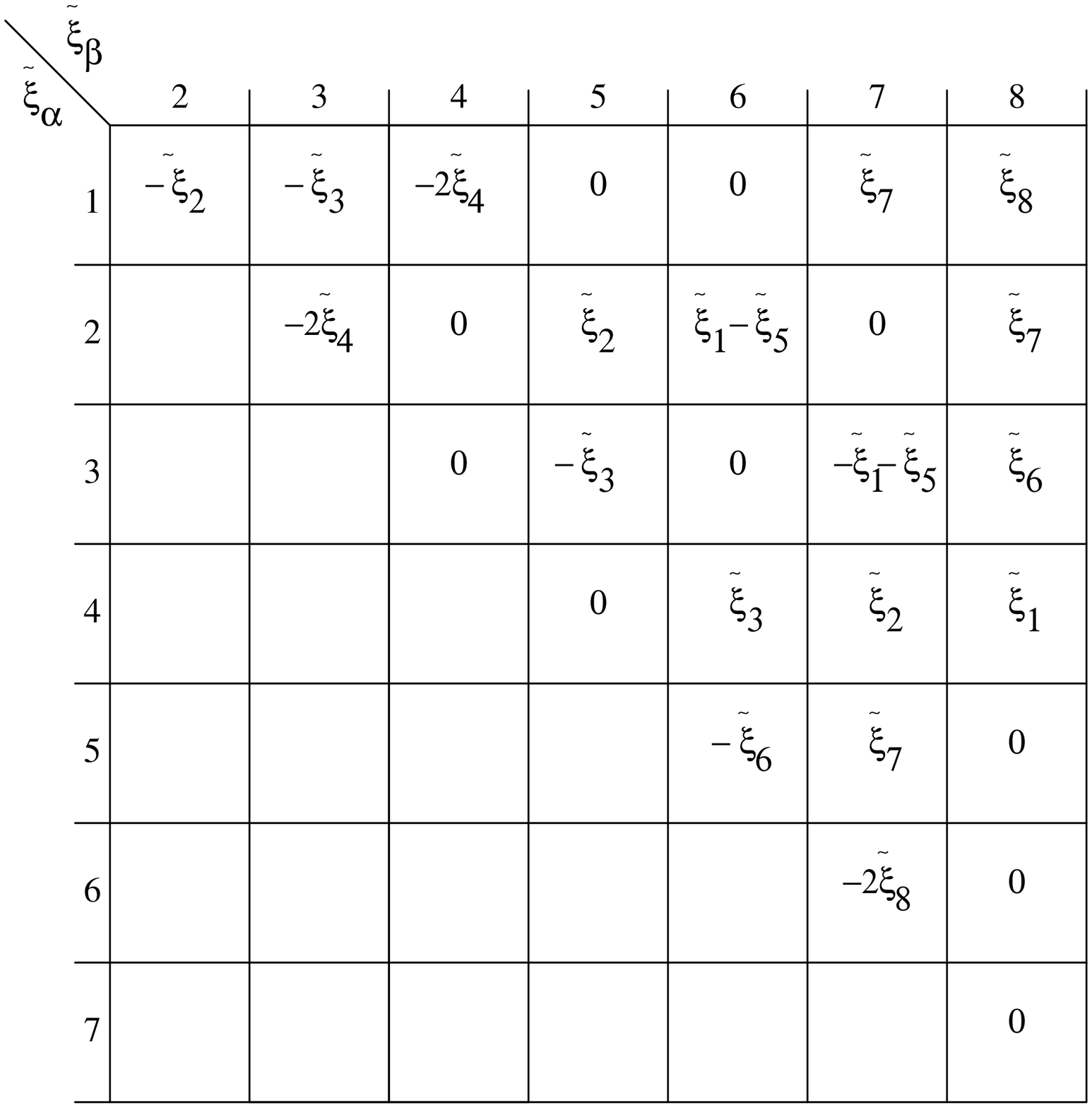,height= 10 cm}}

\end{center}

{\caption{
\baselineskip 3 ex
The commutators $[\tilde{\xi}_\alpha ,\tilde{\xi}_\beta ]$
of the $a^2=3$ Killing vectors.
\label{mult_3}}}
\end{figure}

\end{document}